\documentclass[useAMS,usenatbib,usegraphicx]{mn2e}
\usepackage{url}

\title[The list of tantalum lines for wavelengths calibration of the Hamilton 
echelle-spectrograph]{The list of tantalum lines for wavelengths calibration of
the Hamilton echelle-spectrograph}

\author[Yu.V.Pakhomov]{
Yu.~V.~Pakhomov,$^{1}$
 \\
$^1$Institute of Astronomy, Russian Academy of Sciences, Pyatnitskaya 48,
119017, Moscow, Russia\\
}

\begin{document}
\maketitle

\begin{abstract}
We present solution of the problem of wavelength calibration for Hamilton
Echelle spectrograph using hollow cathode lamp, which was operated at Lick
Observatory  Shane telescope before June~9,~2011. The spectrum of the lamp
claimed to be thorium-argon, contains, in addition to the lines of thorium and
argon, a number of the unrecognized lines identified by us with tantalum. Using
atomic data for measured lines of tantalum and thorium, we  estimated the
temperature of the gas in the lamp as  $T=3120\pm60$~K. From the atomic line
database VALD3 we selected all lines of TaI and TaII which can be seen in the
spectrum of the lamp and compiled a list for the use in the processing of
spectral observations. We note a limitation of the accuracy of calibration due
to the influence of the hyperfine line splitting.
\end{abstract}


\section{INTRODUCTION}

Hollow cathode lamps are the main source of reference spectral lines for
wavelength calibration of the spectra. The lamps themselves are sealed glass
flasks filled by an inert gas, containing inside a cylindrical cathode with
metal sputtering. Under the action of electric current the gas in the lamp is
heated, partially ionized, its atoms become excited and emit photons in the
corresponding wavelengths. In addition, the atoms and ions of the gas begin to
bombard cathode surface, embossing the metal atoms, which, in turn, also emit
photons with  characteristic wavelengths. Resulting spectrum of the lamp
contains the lines of at least two elements, of the  cathode gas and of the
cathode element, in two ionization stages. Most often the gas is argon, helium,
neon or a combination of them, while the cathode is coated by a thin layer of
some metal. The industry produces dozens of lamps with different sputtering of 
cathodes: by aluminum, iron, gold, calcium, chromium, bismuth, cesium, etc. In
the astronomical practice, most common is thorium-argon lamp,  thanks to
numerous spectral lines, more or less evenly distributed over the visible. Often
used are iron-argon lamps, especially for calibration of observations in the
blue spectrum.

Wavelength calibration is an integral part of spectral observations. Poor
calibration is a source of all kinds of errors in the determination of radial
velocities and identification of spectral lines and, hence, it leads to
inaccuracy of scientific results. For accurate calibration, it is necessary to
know the wavelengths of reference wavelines with an error of at least an order
of magnitude lower than the width of the spectrograph instrumental profile
$\delta\lambda\sim0.1\lambda/R$, where $\lambda$ is wavelength, $R$  -- spectral
resolution. For thorium-argon lamps there exist several lists and atlases which
are contained in the well-known systems of observation processing MIDAS and
IRAF. There are also more detailed lists of lines, for instance,
\citep{NOAO_ThAr, 2007A&A...468.1115L, 2014ApJS..211....4R}.

In March 2011, as part of ``A systematic study of NLTE abundance of nearby
dwarfs'' proposal (PI -- Zhao Gang, NAOC, China), the spectra of 6 stars were
taken by the Shane 3~m telescope of Lick Observatory (California, USA) using the
Hamilton Echelle Spectrograph. We used the  CCD e2v CCD203-82 (4k$\times$4k,
$12\mu$, Dewar \#4) and thorium-argon lamp (S\&J Box, Westinghouse WL23418,
symbol ThAr02) as the source of comparison spectrum. Spectral resolution was
$R=60000$. For the processing of observations we applied the package ``echelle''
of the software system MIDAS. We extracted from the CCD-frame 115 echelle orders
in the wavelengths range 3373--10915~\AA. Wavelength calibration begins with
identification of several relatively bright lines of argon and then the code
automatically identifies observed  spectral lines using the list of standard
lines \textit{thar100.tbl}. For high-grade calibration, identification of more
than a half of all observed lines across the spectrum, as well as at least
5~lines in each echelle order are necessary. However, the program was able to
identify only about $5\%$ of the spectral lines. A careful examination of CCD
frame revealed the presence of about 20 unidentified bright lines with
wavelengths from 4700 to 6500~\AA, which are usually not observed in the spectra
of thorium-argon lamps. The lines of thorium were, on average, weakened and, at
the same time, some lines from the list were even not observed in the lamp
spectrum. These two occasions resulted in the failure of wavelength calibration.

Above-described situation forced us to look for the solution of the problem. It
was found by identification of the unrecognized lines with these of tantalum, as
described in Section~2 of the article. For further calibration it was necessary
to compile a list of the lines of tantalum (Section~3). In Section~4 we present
an example of calibration using above-mentioned list of lines and discusses its
accuracy. The results obtained in this study will be useful for high-quality
processing of spectral data obtained by the Hamilton spectrograph using the lamp
under scrutiny before June~9, 2011. Furthermore, a similar pattern of the
location of bright lines of tantalum is observed for another Lick Observatory
lamp (symbol ThAr07, Westinghouse Box, Westinghouse WL32809) that was in
operation from February~17, 1995 to June~19, 2011) and which is, highly-likely,
a ThTaAr lamp also.

Earlier, we  encountered already a similar problem with calibration lamps in
Lick observatory \citep{2013AJ....146...97P}. Therefore, we shall apply a
similar method to solve above-mentioned problem too.   

\section{IDENTIFICATION OF UNKNOWN LINES}

Note, in practice, there are no two identical lamps with hollow cathode.
Differences in the physical characteristics and conditions, such as the pressure
and gas temperature, the voltage, service time, lead to different relative
intensities of spectral lines. However, the strongest lines persist and reflect
a characteristic pattern in the CCD frame. An example of this are very strong
lines of argon in the  7000--8000~\AA\ range. The website of Lick
Observatory\footnote{http://mtham.ucolick.org/techdocs/techref/ThAr/} provides a
possibility to compare the spectra of lamps used in the Hamilton spectrograph.
It is easy to see that most of the dozen lamps show almost identical patterns,
with exception of ThAr01, ThAr02, and ThAr07 lamps. The first of them was
identified by us as a TiAr lamp by \citet{2013AJ....146...97P}.

For identification of unknown lines we carried out a preliminary calibration
using the lines of argon and thorium. For some echelle orders with a sufficient
number of these lines, calibration accuracy better than 0.005~\AA\ was achieved.
For other orders containing only few lines of argon and too weak thorium lines,
we used a two-dimensional solution of the echelle equation, which provided the
accuracy of about 0.03--0.05~\AA. Altogether, 1815 lines were found in the
spectrum of the calibration lamp.  Using the database of the atomic parameters
of spectral lines VALD3 \citep{2015PhyS...90k4005R}, for the vicinity of every
of these lines, for the distances less than  $\Delta\lambda=0.1$~\AA\ (which we
call below detection domain), we constructed a sample of all spectral lines
belonging, possibly, to more than 70 neutral atoms and ions in the first
ionization stage. For each element the total number of possible identifications
was found. Next, detection domain was narrowed to $\Delta\lambda = 0.05$~\AA\
and the number of identifications of the elements was found again. The idea of
the method is that the elements that appear in the detection domain by accident
would show a two-fold reduction in the number of identifications with transition
of $\Delta\lambda$ from 0.1 to 0.05~\AA, while the number of the elements
belonging to the lamp will remain the same or slightly reduce. We deleted from
the list the elements with the number of identifications less than 100
(about~$5\%$), since the elements of cathode usually produce much more spectral
lines.

\begin{figure}
\centering
\resizebox{\hsize}{!}{\includegraphics[clip]{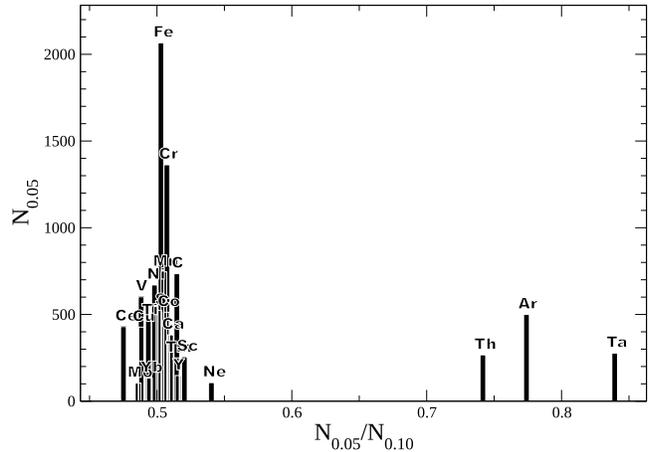}}
\caption{The number of elements that have spectral lines in the detection domain
with the size 0.05~\AA, and relative decrease of this number with decrease of
the size of the domain from 0.10 to 0.05~\AA. \hfill} 
\label{fig:N}
\end{figure}

\begin{figure*}
\centering
\resizebox{\hsize}{!}{\includegraphics[clip]{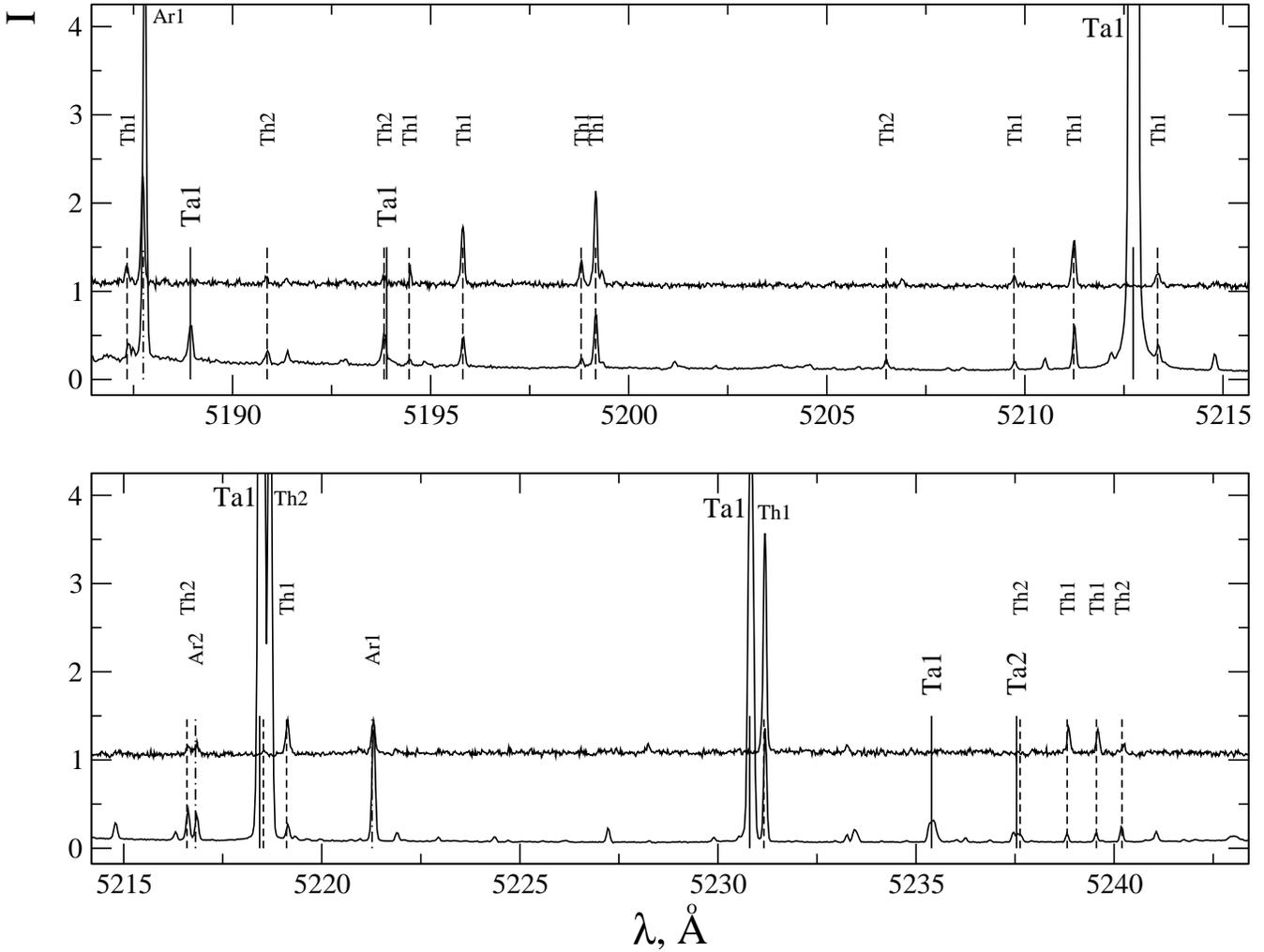}}
\caption{Comparison of the spectrum of the lamp under study with the spectrum of
a typical ThAr lamp installed in the spectrograph EMMI (for visibility, the
spectrum is risen). Thorium and argon lines are shown by dashed lines, while the
lines of tantalum are shown by solid lines.  \hfill} 
\label{fig:iden}
\end{figure*}

Figure~\ref{fig:N} shows the number of identifications $N_{0.05}$ for
$\Delta\lambda=0.05$~\AA\ and its relative change $N_{0.05}/N_{0.10}$. It is
evident that the overwhelming majority of elements is concentrated around
$N_{0.05}/N_{0.10}=0.5$, while three elements (thorium, argon, and tantalum)
have significantly different positions and the number of their identifications
reduces by $15{-}25\%$.

Figure~\ref{fig:iden} shows the comparison of the spectrum of 109th
echelle-order containing three unknown bright lines with the spectrum of the
thorium-argon lamp installed in the spectrograph EMMI (telescope NTT, ESO).
Common details in the spectra are the lines of thorium and argon, while three
bright lines ($\lambda\lambda$~5212.73, 5218.43, and 5230.80~\AA) and a number
of relatively weak lines are present only in the spectrum of the tested lamp.
Checking of the positions of the spectral lines of tantalum confirmed in full
that the cathode elements are tantalum and thorium.

\section{COMPILATION OF THE TANTALUM LINES LIST}

In the approximation of local thermodynamic equilibrium (LTE) without account of
the self-absorption the intensity of a spectral line, produced by spontaneous
emission, is described by expression $I\propto (gf/\lambda^3)\exp(-E/kT)$, where
$g$ is the statistical weight of the upper level, $f$ -- oscillator strength for
given transition, $\lambda$ --  spectral line wavelength, $E$ -- the energy of
excitation of the upper level, $k$ -- Boltzmann constant, and $T$ -- gas
temperature. Strictly speaking, in hollow-cathode lamps LTE-conditions are not
fulfiled and our computations provide only estimates.
           
To compile a list of tantalum lines, it is necessary to choose from the database
spectral lines with registered intensity exceeding the minimum of intensity of
the lines in the lamp spectrum. This requires absolute flux calibration of the
spectrum. Furthermore, it is necessary to know the temperature of the gas, which
can be estimated from the spectral lines by using the values of their absolute
or relative intensity and atomic parameters ($F$, $g$, $\lambda$, $E$) from the
relation 
\begin{equation}
\log\frac{gf}{I\lambda^3} = a+bE,
\label{eq:lgI}
\end{equation}
where $a$ is a constant that determines the
zero point of the scale of the oscillator strengths, $b=5040/T$ -- a constant
characterizing population levels in the Boltzmann distribution.

\subsection{Flux calibration}

Flux calibration of the lamp spectrum was performed using the spectrum of the
star HD~103095, taken the same night. Because the latter star is not a
spectrophotometric standard, the energy distribution was calculated using the
synthetic spectrum code \textit{SynthV} \citep{2003IAUS..210P.E49T} and the
program for modeling of stellar atmospheres \textit{ATLAS9}
\citep{1993KurCD..13.....K}. In the calculation of the model atmosphere, the
following parameters were adopted: effective temperature $T_{eff}=5130$~K,
effective gravity $\log g=4.66$, metallicity $\textrm{[Fe/H]} = {-}1.26$,
microturbulence velocity~0.9km/s \citep{Tania}. The star HD~103095 is
sufficiently close (parallax $0.11^{\prime\prime}$), so the interstellar
reddening was not considered. To refuce the fluxes to absolute scale, we applied
the magnitudes in the bands $U$, $B$, $V$, $R$, $I$, $J$, $H$, $K$, recalculated
using  the relations of \citet{1998A&A...333..231B}. Synthetic spectrum in
absolute fluxes with account for radial velocity of the star and  orbital
velocity of the Earth (in total, ${-}91.5$~km/s) was converted into the MIDAS
format to construct the response function of the spectrograph--CCD system that
allowed to restore the energy distribution in the spectrum of calibration lamp.
The values thus obtained are relative, because it is not known what fraction of
the lamp radiation is intercepted by the spectrograph.

\begin{figure}
\centering
\resizebox{\hsize}{!}{\includegraphics[clip]{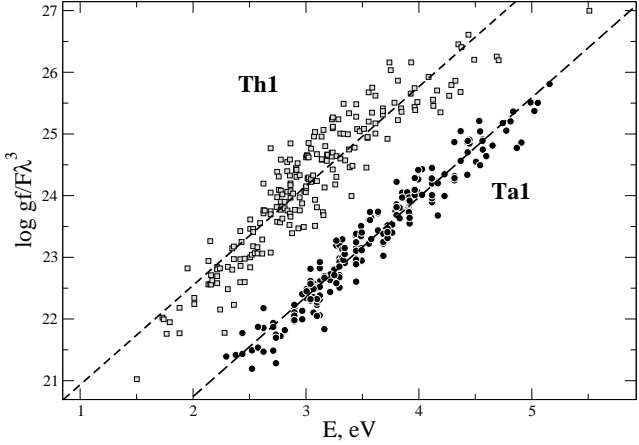}}
\caption{The estimate of gas temperature from the slope of Boltzmann
dependence for the data on TaI and ThI atoms. \hfill} 
\label{fig:I-E}
\end{figure}

\subsection{The estimate of gas temperature}

For all found nonblended lines of TaI (201 lines) and ThI (211 lines) we
measured intensity and obtained atomic parameters from the database VALD3
\citep{2015PhyS...90k4005R}. We used only spectral lines with saturation
threshold not exceeding the threshold of the used CCD. Figure~\ref{fig:I-E}
shows dependence of the left side of Eq.~\ref{eq:lgI} on the excitation
potential of the upper level $E$. The scatter of points in the vertical
direction is, primarily, due to the deviation from LTE and possible errors in
accounting of the scattered light of echelle spectrograph. It can be seen that
tantalum lines form an explicit dependence (correlation coefficient 0.97),
parallel to thorium lines, suggesting similar temperatures, correct
identification of unknown lines and the possibility of evaluation of the
temperature in the LTE approximation. Solution of Eq.~\ref{eq:lgI} for tantalum
-- $a=17.5\pm0.1$, $b=1.16\pm0.03$, for thorium -- $a=19.3\pm0.1$,
$b=1.16\pm0.05$. Hence, the obtained values of the gas temperature are,
respectively, $3120\pm60$~K and $3130\pm90$~K. The resulting temperature is
close to the value of $3200\pm150$~K, found by \citet{2013AJ....146...97P} for
the other lamp of the  Hamilton spectrograph, as it is expected because the
lamps are operated at the same conditions.

\subsection{Tantalum line list}

We have found that the minimum flux recorded in the spectral line of the lamp
under study is $2\times10^{-14}$~erg/cm$^2$. Therefore, to search for all lines
of tantalum, which can be seen in the spectrum of the lamp, we selected from the
database VALD3 all lines satisfying the condition:
$$
I =\frac{gf}{\lambda^3} 10^{{-}17.51-1.616 E}>2\times 10^{-14}.
$$
Also, 10 lines of TaII were identified in the spectrum. The analysis of
intensities of the lines of the ions of tantalum in LTE approximation made it
possible to estimate the concentration ratio:
$$
N\textrm{(TaII)}/N\textrm{(TaI)}\approx0.2
$$

Altogether, 557 TaI lines and 46 TaII lines in the range 3000--12000~\AA\ were
selected from the VALD9 database. As our calculations were performed assuming
LTE, to refine the list, each line was tested for the presence in the spectrum
and the possibility of its use. In the final list, there remained 262 lines of
TaI and 3 lines of TaII between 3387 and 8415~\AA. A part of the list of lines
is shown in the Table~\ref{tab:list}; complete list is available in electronic
form\footnote{ftp://cdsarc.u-strasbg.fr/pub/cats/J/AZh}.

\begin{figure}
\centering
\resizebox{\hsize}{!}{\includegraphics[clip]{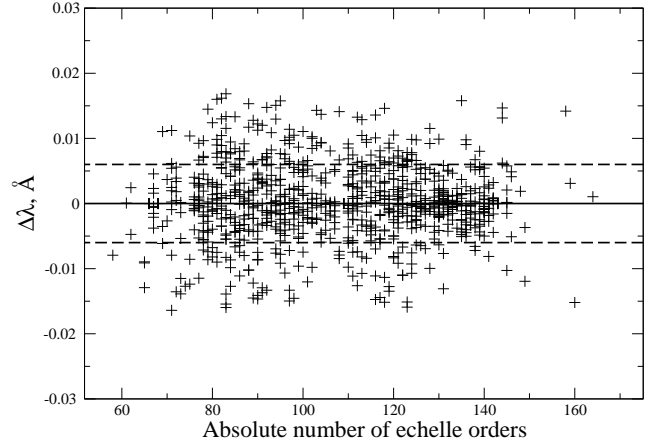}}
\caption{Deviations of the centers of measured lines from the table values for
different echelle orders. Dashed lines limit the region of mean square 
deviation 0.006~\AA.  \hfill} 
\label{fig:calib}
\end{figure}

\begin{table}  
\caption{The list of tantalum lines for wavelengths calibration
(the full list is available electronically)
}
\label{tab:list}
\centering
\begin{tabular}{cc}
\hline
$\lambda$, \AA &  Ion  \\
\hline
...      & ... \\
4205.885 & Ta I \\
4228.622 & Ta I \\
4245.344 & Ta I \\
4279.052 & Ta I \\
4294.380 & Ta I \\
4314.522 & Ta I \\
4318.813 & Ta I \\
4360.792 & Ta I \\
4374.175 & Ta I \\
4378.810 & Ta I \\
4381.885 & Ta I \\
4386.060 & Ta I \\
4398.444 & Ta I \\
...      & ... \\ 
\hline
\end{tabular}
\end{table}

\section{DISCUSSION}

Using the list of tantalum lines together with the list of lines of thorium and
argon, we calibrated the echelle spectrograph Hamilton. Figure~\ref{fig:calib}
shows the deviations of the centers of the measured lines from the Table values
for different echelle orders. The mean square error of calibration of entire
spectrum comprised 0.006~\AA, like in the case of titanium-argon lamp
\citep{2013AJ....146...97P}, but for the range 4000--7000~\AA\ the accuracy was
higher -- up to 0.002--0.003~\AA, while now it is 0.005--0.006~\AA, wherein some
of the lines have deviations exceeding 0.01~\AA. This may be due to the fact
that the tantalum  has an odd number in the periodic table, is subject to the
effect of hyperfine splitting of atomic levels. According to the measurements of
\citet{2003PhyS...68..170M}, the splitting of spectral lines of tantalum
approaches 0.1--0.3~\AA. The uneven distribution of the components leads to the
asymmetry of the measured line and to the shift of its center. For this reason,
the overall accuracy of the calibration becomes worse. The analysis of the
measured lines of tantalum showed that about half of the lines have the width of
about $0.07\pm0.03$~\AA, that generally corresponds to the spectral resolution.
At this, $20\%$ of the lines have the  width of more than 0.2~\AA. Therefore, if
a larger accuracy is needed in the analysis of the spectra, for example, for
measuring the radial velocities with the precision better than 0.5~km/s, the use
of tantalum lines is impractical. For small number of the lines of thorium and
argon in some echelle orders, it is impossible to avoid the use of the lines of
tantalum, but one should consider the real accuracy of the calibration
wavelengths. For the measurements of the line intensity, determination of
abundances of chemical elements and other problems, the accuracy of the
calibration using tantalum lines will be sufficient.

\section{CONCLUSION}

To summarize, in the spectrum of the  hollow cathode calibration lamp, installed
in the spectrograph Hamilton, the presence of unknown lines was revealed. These
lines were identified with tantalum. We determined the temperature of the gas in
the lamp, which allowed  to select from the database of spectral lines VALD3
spectral lines of tantalum, which can be observed in the spectrum of the lamp.
Using compiled  list of lines, we performed wavelength calibration with an
average accuracy of about 0.006~\AA. However, for some lines of tantalum
deviations of the measured positions from the line list values are possible, due
to the effect of the hyperfine splitting that imposes constraints on the
accuracy of calibration for a number of echelle orders.

\section*{Acknowledgments}

This study was partially supported by the Russian Foundation for Basic Research
(project No. 14--02--91153-GFEN\_a) and by the Program of Basic Research of the
Presidium of Russian Academy of Sciences ``Non-stationary phenomena in the
objects of Universe''.

\bibliographystyle{mn2e}
\bibliography{paper}

\end{document}